\documentclass[preprint,amsmath,superscriptaddress,showpacs,showkeys,amssymb,aps,prb]{revtex4-1}
\usepackage{graphicx,multirow,dcolumn,bm}

\begin{document}

%\preprint{APS/123-QED}

\title{Phase separated magnetic ground state in Mn$_3$Ga$_{0.45}$Sn$_{0.55}$C}

\author{E. T. Dias}
\author{K. R. Priolkar}
\affiliation{Department of Physics, Goa University, Taleigao Plateau, Goa 403206 India}
\email{krp@unigoa.ac.in}
\author{A. K. Nigam}
\affiliation{Tata Institute of Fundamental Research, Dr. Homi Bhabha Road, Colaba, Mumbai 400005, India}
\author{R. Singh}
\author{A. Das}
\affiliation{Solid State Physics, Division, Bhabha Atomic Research Centre, Trombay, Mumbai 400085}
\author{ G. Aquilanti}
\affiliation{Elettra-Sincrotrone Trieste S.C.p.A., s.s. 14, km 163.5 I-34149 Basovizza, Trieste, Italy}
\date{\today}

\begin{abstract}
Existence of non-ergodic ground states is considered as a precursor to a first order long range magnetostructural transformation. Mn$_3$Ga$_{0.45}$Sn$_{0.55}$C lies compositionally between two compounds, Mn$_3$GaC and Mn$_3$SnC, undergoing first order magnetic transformation. Mn$_3$Ga$_{0.45}$Sn$_{0.55}$C though crystallizes in single phase cubic structure, exhibits more than one long range magnetic transitions. Using a combination of magnetization, ac susceptibility, neutron diffraction and XAFS techniques it is shown that, though Mn$_3$Ga$_{0.45}$Sn$_{0.55}$C exhibits long range magnetic order, it presents a cluster glassy ground state due to formation of magnetically ordered Ga rich and Sn rich clusters. The clusters are big enough to present signatures of long range magnetic order but are distributed in such way that it limits interactions between two clusters of the same type leading to a frozen magnetic state at low temperatures. The main reason for such a cluster glass state is the difference in local structure of Mn atoms that find themselves in Ga rich and Sn rich clusters.
\end{abstract}


\maketitle

\section{Introduction}
In favoring a ground state with minimum possible energy at absolute zero  the third law of thermodynamics brings about a wide range of disorder to order transitions in systems \cite{Careri1984}. Structural and functional materials exhibiting ordering of physical quantities such as magnetic moment, electric dipole and lattice strain \cite{Xiaobing201090} satisfy thermodynamic requirements, while systems with quenched disorder or frozen randomness undergo non thermodynamic transitions that bring about novel behaviors \cite{Sornette2006}. The widely encountered `glassy' state is known to  rely on the slowing down of kinetics \cite{Jerome1997386, Binder198658, Santen2000405} which challenges the fundamental ``hypothesis of ergodicity" \cite{Binder198658, Mydosh1993}. Within the area of magnetic alloys, spin glasses and cluster spin glasses exemplify the existence of a glassy state in ferromagnetic systems. For instance, even a normal ferromagnet like iron when diluted by non ferromagnetic impurities (e.g. AuFe alloys) undergoes a transformation to a glassy state due to competition between ferromagnetic and antiferromagnetic interactions \cite{Cannella19726}. Similar literature reports on dielectric properties of  BaTiO$ _{3} $ reveal a change from the classical paraelectric to ferroelectric phase transition towards a diffused ferroelectric or relaxor kind as the concentration of the point defects (e.g.  La$ ^{3+} $) is increased \cite{Merz195391, Zhang200487, Petrovic201162}. Governed by the same physical principle of destruction of long-range order due to strong  frozen disorder a different class of glass referred to as `strain glass' (or glassy-R martensite) can be obtained by doping point defects into a normal martensitic alloy. Doping excess nickel ($ \sim $1.5\%) into the widely recognized martensitic NiTi alloy destroys the martensitic transformation by creating a strain glass regime giving rise to a non-martensitic (Ti$ _{48.5} $Ni$ _{51.5} $) alloy that continues to exhibit properties like shape memory effect and superelasticity that otherwise stringently depend on the reversible martensitic transformation \cite{Xiaobing200595, Xiaobing200697, Xiaobing200776}.

Founded in frustration and randomness of isolated moments a study of the functional properties of disordered alloys \cite{Tegus2002319, Lacroix2011, Cong2014251, Liu201266, Sherrington2012, Derzhko201529} grew in the hope of understanding their formation, magnetic ordering and the long range RKKY interactions between them \cite{Cannella19726}. Among the existing geometrically frustrated $3d$ transition metal based antiperovskite compounds displaying a fascinating array of properties ranging from giant magnetoresistance \cite{Kamishima200063, Li200572} to a large magnetocaloric effect (MCE) \cite{Tohei200394, Lewis200393, Aczel201490}, two isostructural compounds, Mn$ _{3} $GaC and Mn$ _{3} $SnC exhibiting first order, volume discontinuous magnetostructural transitions have been extensively studied. Technologically significant properties such as a large inverse table like MCE in Mn$ _{3} $GaC have been attributed to the first order transition  (T $ \simeq $ 160 K) from a FM to an antiferromagnetic (AFM) state described by propagation vector $k = [\frac{1}{2}, \frac{1}{2}, \frac{1}{2}]$ \cite{Fruchart19708, Lewis200393}. Although theoretically the magnetic properties of such antiperovskite compounds originate from strongly hybridized Mn $3d$ and C $2p$ states \cite{Shao2013113, Shao201483, Shim200266, Jardin198346} ample differences have been identified in the magnetic properties exhibited by Mn$_3$SnC. Replacing Sn at the Ga site  not only causes  the first order transition to be expressed at the PM to ferrimagnetic ordering at much higher temperatures ($ \sim $280 K) but also causes the MCE to change to the conventional type with little or no field variation seen in the position of the magnetic entropy peak  \citep{Wang200985, Dias20141, Dias2015117}. Our previous work describing the effect of Sn substitution on magnetic properties of Mn$ _{3} $Ga$ _{1-x} $Sn$ _{x} $C compounds illustrates the evolution of this first order transition while highlighting the coexistence of magnetic phases for $0.41  \leq  x  \leq  0.71$ \citep{Dias20141}. Questions arise as to how the two magnetic phases are supported in one crystal structure with magnetic atoms occupying equivalent lattice sites. Do the two magnetic phases interact with each other and if they do, what could be the most physically accepted ground state of these compounds?

Generally, frustrations in a system leading to non ergodic ground states occur when impurities are added to a compound undergoing a phase transformation as in the case of AuFe alloys or La doped BaTiO$ _{3} $ or Ni rich NiTi alloys. Mn$_3$Ga$_{0.45}$Sn$_{0.55}$C belongs to that region of coexisting magnetic phases in the magnetic phase diagram of Mn$_3$Ga$_{1-x}$Sn$_{x}$C. Although the compound crystallizes with a cubic antiperovskite type structure (\textit{Space group: Pm$\bar 3$m}) that has crystallographically equivalent Mn atoms at the face centers, C at the body center and Ga or Sn atoms occupying the corners of the cube, magnetization results showcase two apparent first order transitions. The first magnetic transition is from an ``enhanced paramagnetic" state\cite{Cakir2013344} to an AFM order at 155 K akin to that in Mn$_{3} $GaC followed by a rapid increase in magnetization to a ferromagnetic state with distinct similarities to results obtained for Mn$_{3}$SnC. Therefore the interest of this article lies in the unprecedented significance of discovering a non ergodic ground state in a solid solution where both the end members, Mn$_3$GaC and Mn$_3$SnC undergo a first order transformation.

\section{Experimental}
The spatial distribution of coexisting phases in polycrystalline Mn$_3$Ga$_{0.45}$Sn$_{0.55}$C has been studied using a combination of various diagnostic tools. Synthesis of the antiperovskite material used for all characterizations was done using the solid state reaction technique wherein a stoichiometric mixture of the starting materials in elemental form (Mn, Ga, Sn and C) along with 15 wt.\% excess carbon powder was mixed and pressed into a pellet that was sealed in an evacuated quartz tube. The sealed tube was initially heated at 1073 K for the first 48 hours and the final annealing was carried out at 1150 K for 120 hours. Field and temperature dependent magnetic characterizations in the 5 K to 390 K temperature range in applied magnetic fields of 0.01 T and 0.5 T and between $\pm$7 T were carried in a Magnetic Property Measurement System (Quantum Design, U.S.A ) using protocols described in Ref [32]. The phase purity and structure of the prepared compound was obtained from room temperature x-ray diffraction pattern recorded using Cu K$ _{\alpha} $ radiation.  Neutron diffraction data at various temperatures were collected on the PD2 neutron powder diffractometer ($ \lambda =1.2443$ \AA~) at Dhruva Reactor (Bhabha Atomic Research Center, Mumbai, India). AC magnetic susceptibility measurements were performed using a Physical Properties Measurement System (Quantum Design, U.S.A ). Measurements in the 2-300 K temperature range were carried out at various excitation frequencies (33 $ \leq $ f $ \leq $ 10000 Hz) by applying an AC amplitude H$ _{ac} $=10 Oe after cooling the sample in zero field \cite{Scholz20166}. The local structure surrounding the Mn$  _{6}$C octahedra in compound  Mn$_3$Ga$_{0.45}$Sn$_{0.55}$C was studied using extended x-ray absorption fine structure (EXAFS) spectra recorded at the XAFS beamline at Elettra, Trieste \cite{Giuliana2009190}. Both incident (I$ _{0} $) and transmitted (I) intensities were simultaneously measured at the Mn K edge (6539 eV) within the -200 to 1300 eV range using an ionization chamber filled with appropriate gases at 300 K (RT) and 80 K (LT).  An optimized value of  the absorption edge jump ($ \Delta\mu $) in transmission mode was obtained by selectively adjusting the number of piled layers of Mn$_3$Ga$_{0.45}$Sn$_{0.55}$C  powder coated scotch tape. Prior to reducing the Mn K-edge data to obtain the EXAFS ($ \chi(k) $) signal using well established procedures in the Demeter program \cite{Ravel200512}, the edge energy was calibrated to be in tune with the first inflection point of Mn metal foil.

\section{Results and Discussion}
\begin{figure}
\begin{center}
\includegraphics[width=\columnwidth]{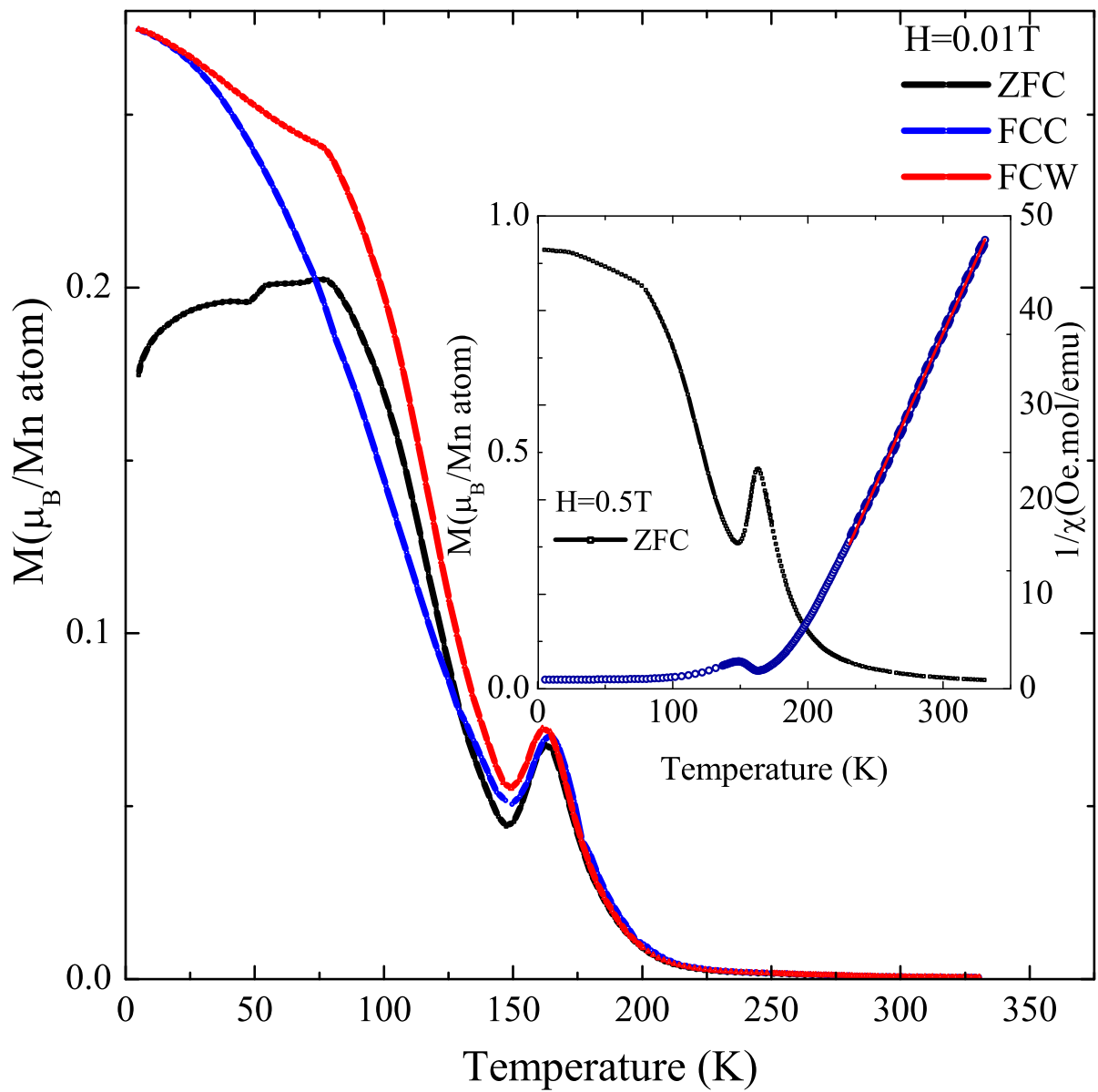}
\caption{Magnetic characterization of Mn$_3$Ga$_{0.45}$Sn$_{0.55}$C between 5 K and 300 K in an applied field of 0.01 T under Zero field cooled (ZFC), Field cooled cooling (FCC) and Field cooled warming (FCW) processes. Inset shows the inverse susceptibility plot to the ZFC data at H=0.5 T.}
\label{fig:mvst}
\end{center}
\end{figure}

Rietveld refined x-ray diffraction pattern confirms the formation of a single antiperovskite phase with minor impurities of C and MnO (each less than 0.5\%). However, magnetization results highlight the coexistence of competing magnetic phases. Recorded in the 5-300 K temperature range with an applied field of  0.01 T the magnetization curve in Figure \ref{fig:mvst} exhibits a transition from an enhanced paramagnetic state to an AFM state at $\sim$155 K followed by a rapid increase in magnetization arising due to ferromagnetic interactions characteristic of Sn rich regions \cite{Dias20141}.

The marked hysteresis observed between the cooling and warming curves establishes the first order nature of both these transitions. Apart from the large splitting observed between ZFC and FC curves a broad maximum is seen in the ZFC curve in the ferromagnetic state ($ \sim $30 K)  below which the magnetization decreases with decreasing temperature. A Curie Weiss fit, as shown in the inset of Figure \ref{fig:mvst}, to high temperature magnetization data recorded in 0.5 T gives a PM Curie temperature $ \theta_{P} $ = 182 K.  Based on this quite large and positive value of $\theta_P$, the observed decrease of magnetization in the ZFC curve cannot be interpreted as a transition to the AFM state \cite{Maletta19789}.

\begin{figure}
\begin{center}
\includegraphics[width=\columnwidth]{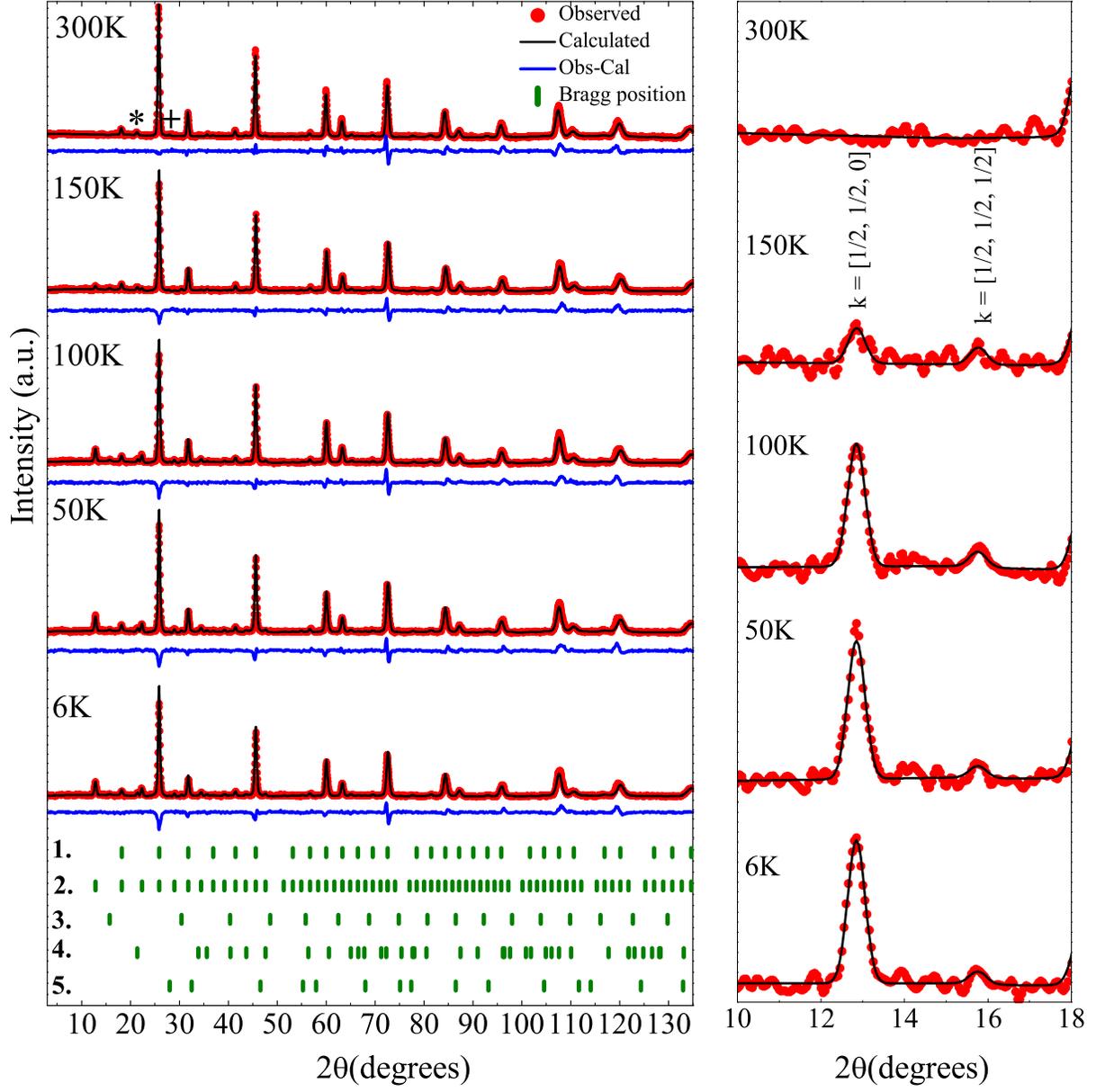}
\caption{a. Series of Rietveld refined neutron diffraction patterns between 6 K and 300 K. Enumerated Bragg positions correspond to 1. Mn$_3$Ga$_{0.45}$Sn$_{0.55}$C (chemical unit cell) 2. and 3. Magnetic reflections of Mn$_3$Ga$_{0.45}$Sn$_{0.55}$C associated with propagation vectors $k = [{1\over 2}, {1\over 2}, 0]$ and $k = [{1\over 2}, {1\over 2}, {1\over 2}]$ 4. Graphite (\lq$\ast$\rq indicates most intense peak at 300 K) and 5. MnO (maximum peak indicated by a \lq$+$ \rq sign at 300 K). b. Angular 2$\theta$ range between 10$^\circ$-18$^\circ$ highlighting the temperature dependent development of magnetic peaks associated with the long range AFM order in Mn$_3$Ga$_{0.45}$Sn$_{0.55}$C.}
\label{fig:neutronpatterns}
\end{center}
\end{figure}

\begin{figure}
\begin{center}
\includegraphics[width=\columnwidth]{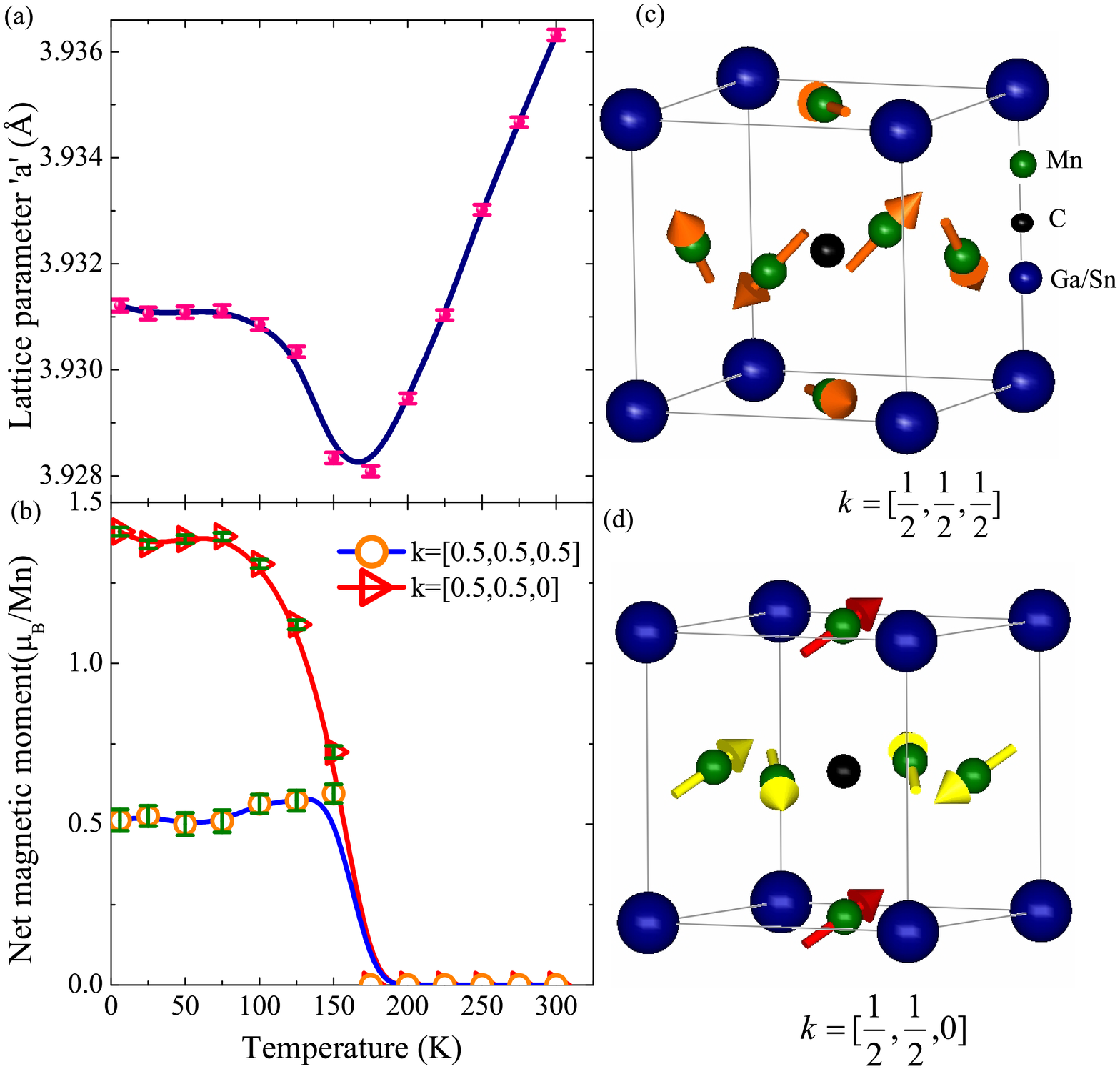}
\caption{a. Mn$_3$Ga$_{0.45}$Sn$_{0.55}$C lattice parameter variation between 6 K and 300 K. b. Thermal evolution of the averaged ordered AFM components. c-d. Three dimensional representations of the magnetic spins associated with the Mn$_3$Ga$_{0.45}$Sn$_{0.55}$C structure as determined from neutron diffraction pattern recorded at 6 K.}
\label{fig:magstr}
\end{center}
\end{figure}

To further explore the complexity of the magnetostructural ordering in this compound, neutron diffraction patterns recorded between 6 K and 300 K in the angular range of 3$^{\circ}$ - 135$ ^{\circ}$ were analyzed.  Using the Rietveld method implemented in the FullProf suite program \cite{Ravel200512}, the crystal structure in the PM state was refined from diffraction data collected at 300 K. Refinement results presented in Figure \ref{fig:neutronpatterns}a, are consistent with the Rietveld refined room temperature XRD pattern in Ref [\onlinecite{Dias20141}]. Occupancy parameters of Mn, Ga, Sn and C atoms obtained through iterative refinements of neutron and x-ray diffraction patterns \cite{Amos200273} (Mn = 2.95$ \pm $0.03; Ga = 0.49$ \pm $0.04; Sn = 0.51$ \pm $0.04 and C = 1.01$ \pm $0.01) are well within the error range. In particular, the occupancy parameters of C and Mn were obtained from refinement of neutron diffraction patterns while those of Ga and Sn were obtained from refinement of x-ray diffraction patterns. The crystal structure is cubic at all temperatures but the thermal evolution of lattice parameters plotted in Figure \ref{fig:magstr}a shows a discontinuity at about 155 K which coincides with enhanced paramagnetic to AFM transition.  At 300 K, the lattice constant $a$ = 3.93632(9) \AA~ obtained for Mn$_3$Ga$_{0.45}$Sn$_{0.55}$C lies between that of Mn$_3$GaC and Mn$_3$SnC \cite{Dias20141}. With decreasing temperature the lattice parameter decreases almost linearly down to 175 K before undergoing an abrupt increase of about $ 0.3\%$ at T $\simeq$ 150 K. Beyond this the unit cell parameter remains almost constant down to 6 K.

A closer look at Figure \ref{fig:neutronpatterns} points to the fact that as temperature decreases additional reflections that cannot be described solely on the basis of the chemical unit cell begin to appear for T$ \leq $150 K. Contemplating the magnetization results and the temperatures at which these new peaks begin to appear one can assume them to be of magnetic origin. A search for a wave vector that yields intensities corresponding to this set of  magnetic reflections indicated that the extra peaks could only be indexed assuming two magnetic propagation vectors $k = [\frac{1}{2}, \frac{1}{2}, \frac{1}{2}]$ and $k = [{1\over 2}, {1\over 2}, 0]$ that order independently of each other. Figure \ref{fig:neutronpatterns}b featuring a magnified view of the 10$ ^{\circ}$-18$ ^{\circ}$ $2\theta$ range highlights the magnetic peaks associated with these two vectors.

Fixing the magnetic reflections to an absolute scale determined by nuclear reflections in the chemical unit cell at 300 K, the thermal evolution of magnetic moments was studied in detail. The arrows in Figures \ref{fig:magstr}c and \ref{fig:magstr}d provide a graphic illustration of the magnetic coupling observed between ordered Mn moments in the cubic unit cell of compound Mn$_3$Ga$_{0.45}$Sn$_{0.55}$C at 6 K. As portrayed by Figure \ref{fig:magstr}c, Mn spins in the (1 1 1)  planes have collinear alignment, however, two such consecutive (1 1 1) planes couple antiferromagnetically giving rise to a doubled unit cell associated with the $[\frac{1}{2}, \frac{1}{2}, \frac{1}{2}]$ propagation vector. The magnetic moment per Mn atom was determined to be 0.59$ \pm $0.03$ \mu_{B} $ as compared to 1.54$ \mu_{B} $ obtained for Mn$ _{3} $GaC at 150 K\cite{Cakir2014115}.
The stronger magnetic reflection at $2\theta = 12.8^{\circ}$ indexed using the $[\frac{1}{2}, \frac{1}{2}, 0]$ wave vector generates the spin alignment exemplified in Figure \ref{fig:magstr}d. The unit cell parameters for the resulting magnetic model are $a \sqrt{2} $, $a \sqrt{2} $, $a$; where $a$ is the lattice parameter of the chemical unit cell. Although the Mn spins confined to the a-b plane are coupled antiferromagnetically, a small canting along the  $[\frac{1}{2}, \frac{1}{2}, 0]$ direction gives rise to a net ferromagnetic moment ($0.37 \pm 0.2 \mu_B$) and is responsible for the rapid rise of magnetization below 150 K. Temperature variation of magnetic moment plotted in Figure \ref{fig:magstr}b indicates that both the magnetic phases have the same ordering temperature of about 150 K which coincides with the volume expansion of the chemical unit cell (\ref{fig:magstr}a).

Overwhelming experimental evidence of existence of two magnetically ordered phases, both antiferromagnetic but one with a finite net magnetic moment over a substantial temperature regime naturally spurs interest in the phase separated nature of the compound. The questions are, whether the two magnetic phases interact and does this interaction results in any kind of frozen order. This is important due to differences in behavior of ZFC and FC magnetization at low temperatures. Characteristic signatures of a glassy behavior include (a) deviation of the ZFC curve from the FC curves \cite{Wu20065}, (b) frequency dependence of low field AC susceptibility, (c) presence of disordered structure and (d)  local or short range order \cite{Xiaobing201090}.  In the present case, despite the fact that the broad maximum observed in the ZFC curve in Figure \ref{fig:mvst} is typical for both spin glass and cluster glass materials, the continuous increase in magnetization with decreasing temperature as observed in the field cooled curves of Mn$_3$Ga$_{0.45}$Sn$_{0.55}$C suggests cluster glass to be more probable over a spin glass state.

\begin{figure}
\begin{center}
\includegraphics[width=\columnwidth]{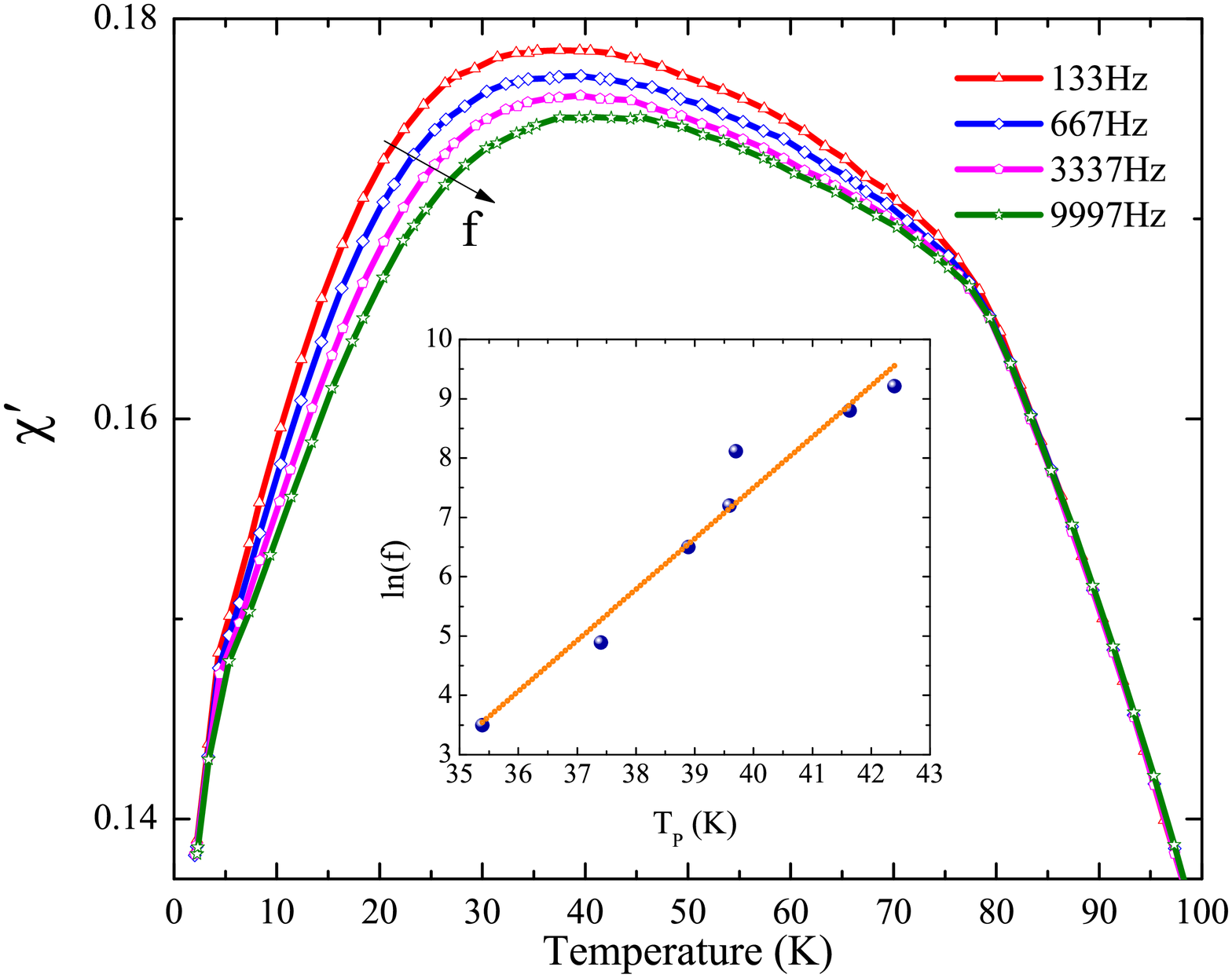}
\caption{Temperature dependence of the real part of AC susceptibility measured for Mn$_3$Ga$_{0.45}$Sn$_{0.55}$C between 2 K to 100 K at different frequencies. Inset shows variation of the peak temperature T$  _{P}$ with the frequency of the AC field in a Vogel-Fulcher plot.}
\label{fig:susceptibility}
\end{center}
\end{figure}

To further confirm the glassy nature of the ground state of this compound, AC susceptibility was measured at frequencies varying between 33 to 9997 Hz across the entire temperature range with an AC driving field of 10 Oe. Temperature dependence of the real ($ \chi' $) component of AC susceptibility displayed in the main panel of Figure \ref{fig:susceptibility} summarizes the results of AC magnetic susceptibility between 2 K and 100 K. The observed maximum $T_{f}$ with a clear frequency dependence  i.e. a systematic shift towards higher temperatures with decreasing amplitudes observed for increasing frequencies is characteristic of magnetic glassy transitions.  A logarithmic frequency dependence of the peak temperature defined as a maximum on the $ \chi' $ curve follows the Vogel Fulcher empirical law fairly well as shown in the inset of Figure \ref{fig:susceptibility} indicative of a \textit{glassy} ground state with frozen magnetic moments. The relative shift of spin freezing temperature ($\delta T_{f}$ = $ \bigtriangleup T_{f} $/T$ _{f} $ $ \bigtriangleup log\nu $, $ \bigtriangleup $ represents a change in the corresponding quantity)  estimated as 0.0797 lies between the values reported for canonical spin glasses (AuMn $\delta T_{f}$ = 0.0045) \cite{Mulder198225} and superparaamagnetic systems ($ \alpha$-(Ho$  _{2}$O$  _{2}$(B$  _{2}$O$  _{3}$))  $\delta T_{f}$ = 0.28) \cite{Mydosh1993} and is close to that of typical cluster glass systems \cite{Mahajan201426}.

\begin{figure}
\begin{center}
\includegraphics[width=\columnwidth]{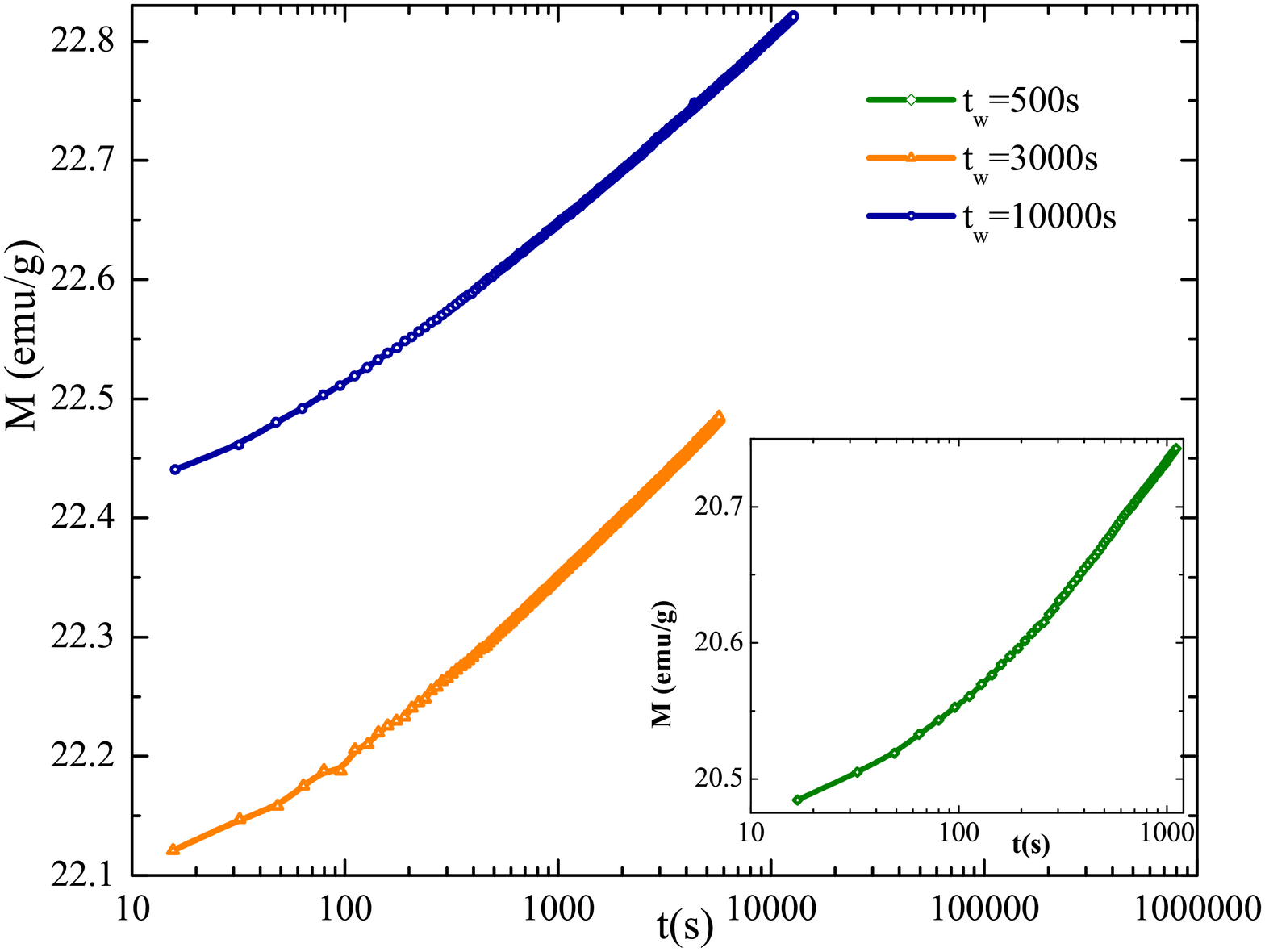}
\caption{Magnetization (M) as a function of time (t) at 10 K recorded under a magnetic field of 0.01 T after the sample was cooled in ZFC mode followed by a subsequent waiting time of $t_{w}$=500s, 3000s and 10000s.}
\label{fig:timedependent}
\end{center}
\end{figure}

These results imply that the compound undergoes one or more relaxation processes with characteristic relaxation time constants. A study of the time dependent response of  magnetic spins in ZFC magnetization was carried out using the following protocol. The sample was first cooled from 300 K to the measurement temperature $\sim $10 K in zero applied field. A waiting time $t_{w}$ was used to equilibrate the spin system before applying a magnetic field of 0.01 T and recording magnetization (M) as a function of observation time `t'\cite{Nigam200266, Cong2014251, Scholz20166}. Time dependent response of the magnetic spins in the present compound to the external applied field is characterized by a distribution of two or more relaxation times as indicated by the exponential growth of M versus t variation in Figure \ref{fig:timedependent}.

Neutron diffraction studies indicate presence of two long range antiferromagnetic orders coexisting below 150 K. These two antiferromagnetic orders are similar to those in the two end members Mn$_3$GaC and Mn$_3$SnC, therefore pointing towards a magnetic phase separation into Ga rich and Sn rich regions. Formation of such clusters is also supported by ac susceptibility studies wherein the peak in $\chi'$ follows Vogel-Fulcher law as a function of frequency. However, structurally the compound crystallizes in a single phase. This is despite the fact that the unit cell of Mn$_3$SnC is larger ($\sim$ 3\%) than that of Mn$_3$GaC. This points towards presence of local structural distortions, especially around Mn atoms.

\begin{figure}
\begin{center}
\includegraphics[width=\columnwidth]{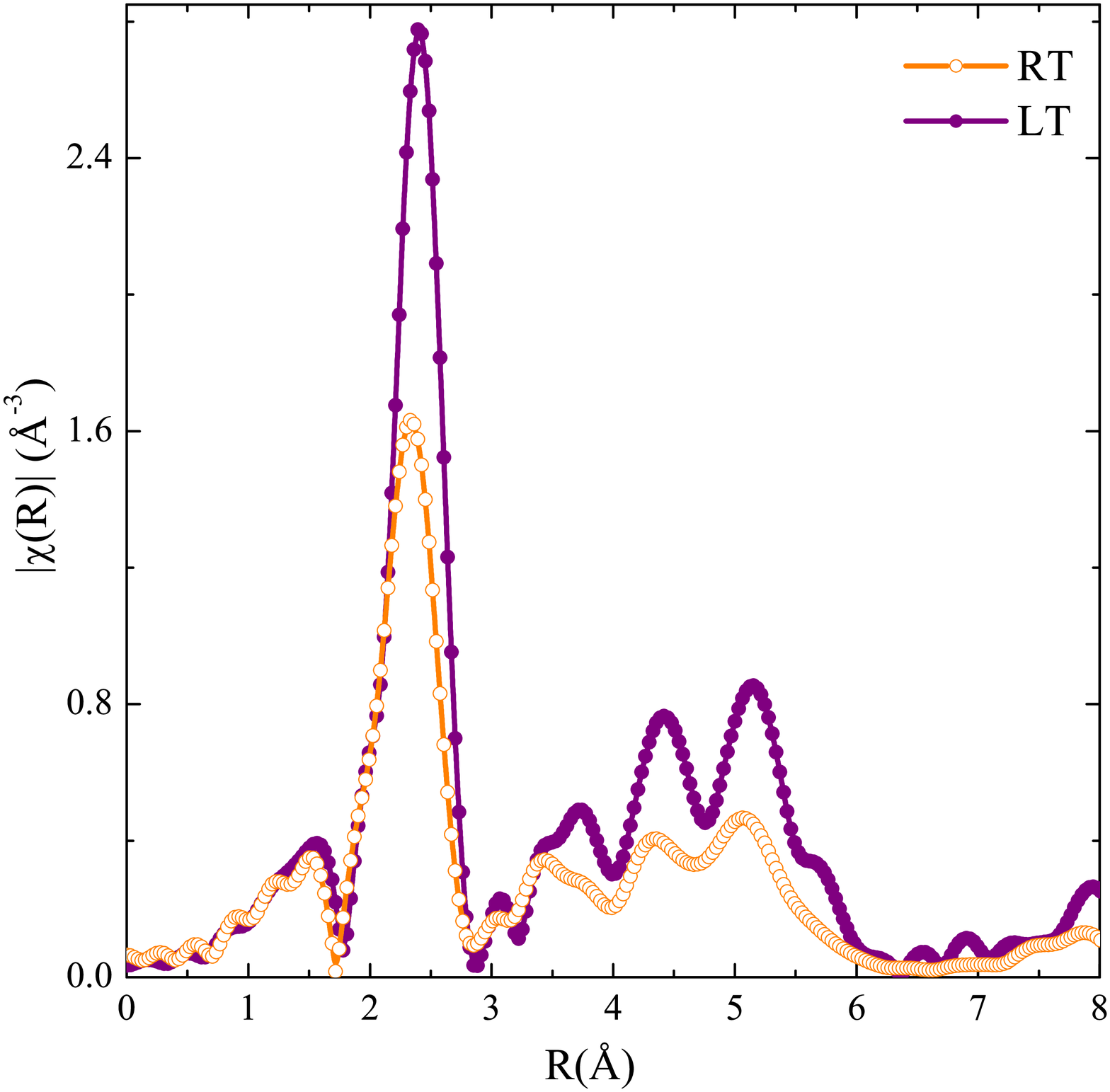}
\caption{Magnitude of Fourier transformed EXAFS spectra obtained at the Mn K-edge in Mn$_3$Ga$_{0.45}$Sn$_{0.55}$C at room temperature (RT) and liquid nitrogen temperature (LT).}
\label{fig:exafs}
\end{center}
\end{figure}

To probe for such possible distortions in the local structure of the Mn that could explain the coexistence of two magnetic phases EXAFS data recorded at the Mn K edge in Mn$_3$Ga$_{0.45}$Sn$_{0.55}$C at RT and LT were analyzed. Magnitudes of the Fourier transform (FT) of both sets of EXAFS data $ \chi (k) $, extracted over a $k$ range of  3.0 - 14.0 \AA$^{-1} $ exhibit two peaks in the R range of 0 \AA~ to 3 \AA~ as shown in Figure \ref{fig:exafs}. The first peak results from scattering by the two nearest neighbor C atoms while the second and the main peak around 2.5 \AA~ includes contributions from the next nearest neighbors i.e. Mn, Ga and Sn atoms. Presence of local structural distortions can also be seen from the the RT and LT plots in Figure \ref{fig:exafs}. With decrease in temperature, all peaks grow in intensity due to decrease in mean square disorder in atomic positions but the peak due to Mn-C correlation exhibits only a small increase. This is a clear indication of presence of local structural disorder.

EXAFS spectra were fitted to two different structural models. The first one was based on the cubic structure of the compounds (cubic model) and all the bond distances were allowed  to contract/expand by  the same factor. The resulting best fit calculation in R-space shows a reasonably good agreement with the experimental data.  In particular, we found that the bonds are contracted by 1\% with respect to the starting structural cubic model (see Table \ref{Table1}).

\begin{table}[]
\caption{Local structural parameters obtained from Mn K edge EXAFS analysis recorded at room temperature in Mn$_3$Ga$_{0.45}$Sn$_{0.55}$C.Here C.N. is the coordination number, R$_{300K}$ is the bond distance calculated from neutron diffraction data recorded at 300K, R denotes the bond distances obtained from EXAFS analysis and $\sigma^2$ denotes the mean square disorder in R obtained from EXAFS analysis}
\label{Table1}
\resizebox{\columnwidth}{!}{%
\begin{tabular}{ccccccc}
\\
\hline
\multirow{2}{*}{Bond} & \multirow{2}{*}{C.N.} & \multirow{2}{*}{R$ _{300K}$(\AA)} & \multicolumn{2}{c}{cubic model} & \multicolumn{2}{c}{delr model} \\
 &  &  & R(\AA) & $ \sigma^{2} $ & R(\AA) & $ \sigma^{2} $ \\
\hline
\hline
Mn-C & 2 & 1.968 & 1.948(2) &0.002(1)& 1.953(12) & 0.004(1) \\
Mn-Mn & 8 & 2.783 & 2.755(2) &0.009(1)& 2.748(27) & 0.006(1) \\
Mn-Ga & \multirow{2}{*}{4} & 2.783 & 2.755(2) &0.003(2)& 2.827(48) & 0.00001(900) \\
Mn-Sn &  & 2.783 & 2.755(2) &0.005(3)& 2.806(43) & 0.001(10) \\
Mn-C-Mn & 16 & 3.359 & 3.326(2) &0.02(5)& 3.146(68) & 0.003(10) \\
Mn-Mn2 & 6 & 3.936 & 3.896(2) &0.014(4)& 3.829(37) & 0.009(3) \\
Mn-C-Mn2 & 4 & 3.936 & 3.896(2) &0.009(3)& 3.868(30) & 0.002(2) \\
Mn-C-Mn2-C & 2 & 3.936 & 3.896(2) &0.004(2)& 3.907(24) & 0.0006(10) \\
R-factor & & & 0.0359 &  & 0.0057 & \\
\hline
\end{tabular}
}
\end{table}

In order to get a better agreement by accounting of atomic size differences between Mn, Ga and Sn, in the second model the structural restrictions were relaxed and the individual correlations were allowed to vary independently (delr model). Though, as in previous case the fit was reasonably good, the obtained Mn-Sn bond distance was shorter than Mn-Ga bond distance (see Table \ref{Table1}). This is impossible as Sn atom being larger than Ga atom, Mn-Sn bond distance would be longer than Mn-Ga distance. Further more, this model also resulted in very low values of $\sigma^2$ especially for Mn-Ga and Mn-C-Mn-C focussed multiple scattering path at 3.91 \AA. Due to increase in number of free parameters, the statistical error bars obtained from fitting were also quite large. Thus despite a very low R-factor, this model is less reliable.

\begin{figure}
\begin{center}
\includegraphics[width=\columnwidth]{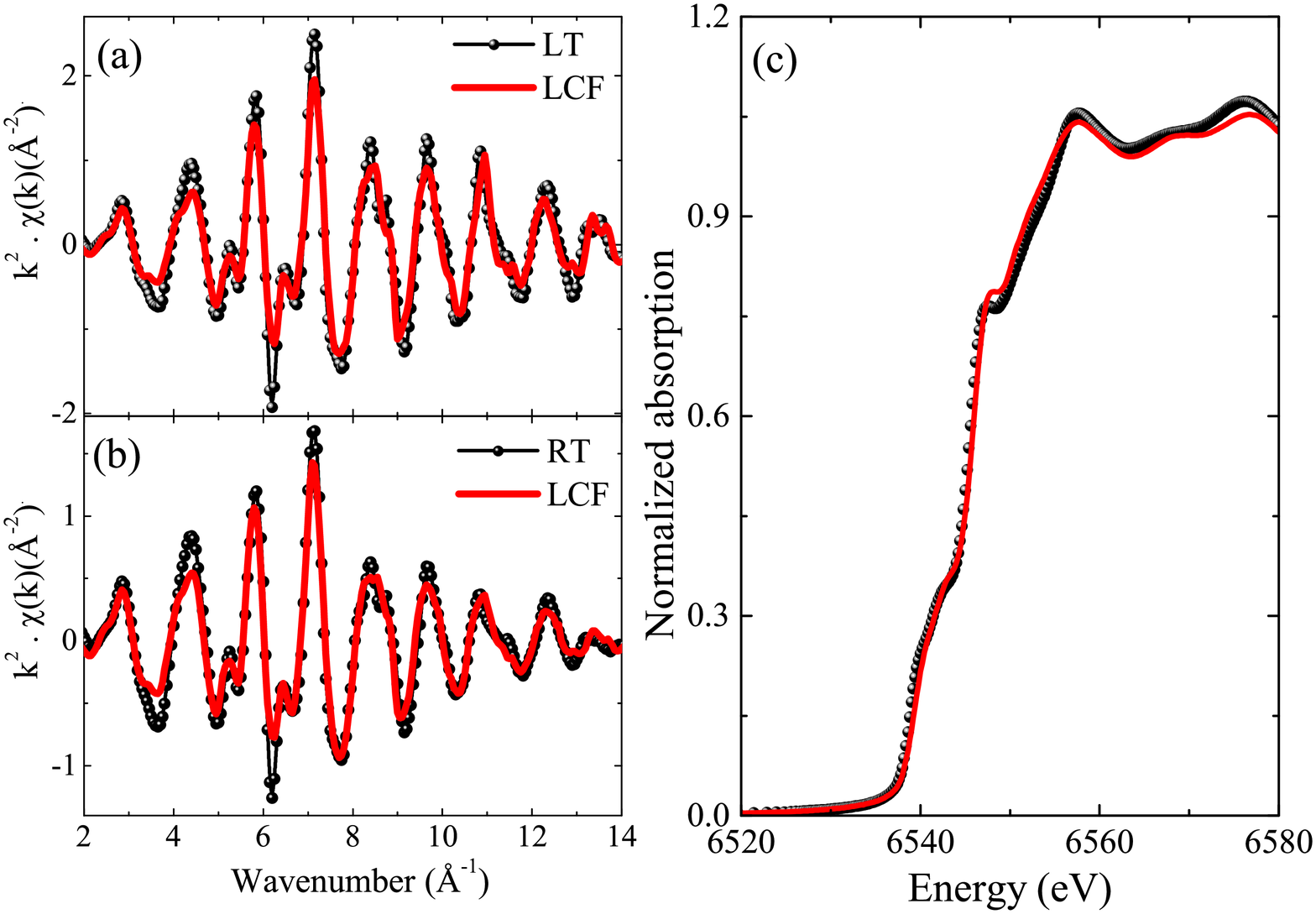}
\caption{The $ k^{2} $-weighted EXAFS spectra of Mn$_3$Ga$_{0.45}$Sn$_{0.55}$C recorded at (a) Low  temperature Mn K-edge and (b) Room temperature Mn K-edge and the corresponding linear combination fit (solid line). (c) LCF analysis to XANES region (-20 eV to +30 eV with respect to absorption edge) of Mn K edge in Mn$_3$Ga$_{0.45}$Sn$_{0.55}$C.}
\label{fig:lcf}
\end{center}
\end{figure}

The above analysis indicates that it would be difficult to map the local disorder without any prior knowledge of at least one or two among Mn-Ga, Mn-Mn and Mn-Sn bond distances. To get a sense of the local structure around Mn we tried a different approach of using a linear combination fitting (LCF) with the experimental spectra of Mn$ _{3} $GaC and Mn$ _{3} $SnC at respective temperatures as standard components to describe the EXAFS spectra of Mn$_3$Ga$_{0.45}$Sn$_{0.55}$C at RT and LT. This has been done using the LCF routine of the Athena software package [\onlinecite{Ravel200512}]. Constraints such as reference spectra having no negative components and sum of coefficients normalizing to unity were imposed while fitting the data in $k$ space in the range 3.0 - 14.0 \AA$^{-1}$ and $k$ weights of 1, 2 and 3. The $k^2$ weighted Mn K edge EXAFS data along with the linear combination fit at both RT and LT are given in Figure \ref{fig:lcf}. The compositions obtained were about 60$\pm$3\% of Mn$_3$GaC and 40$\pm$3\% of Mn$_3$SnC. This hinted at a possibility of Mn atoms having two different local structures. Those Mn atoms that find themselves in Ga rich environment could have local structure similar to that in Mn$_3$GaC while those with Sn rich environment around them may find themselves in a local structural environment similar to Mn$_3$SnC. LCF analysis of the XANES region (-20 eV to +30 eV with respect to the edge energy) using $\Delta$E$_0$ values obtained in EXAFS analysis also gave similar fractional compositions at 53$\pm$4\% of Mn$_3$GaC and 47$\pm$4\% of Mn$_3$SnC at LT and is presented in Figure \ref{fig:lcf}(c).

\begin{figure}
\begin{center}
\includegraphics[width=\columnwidth]{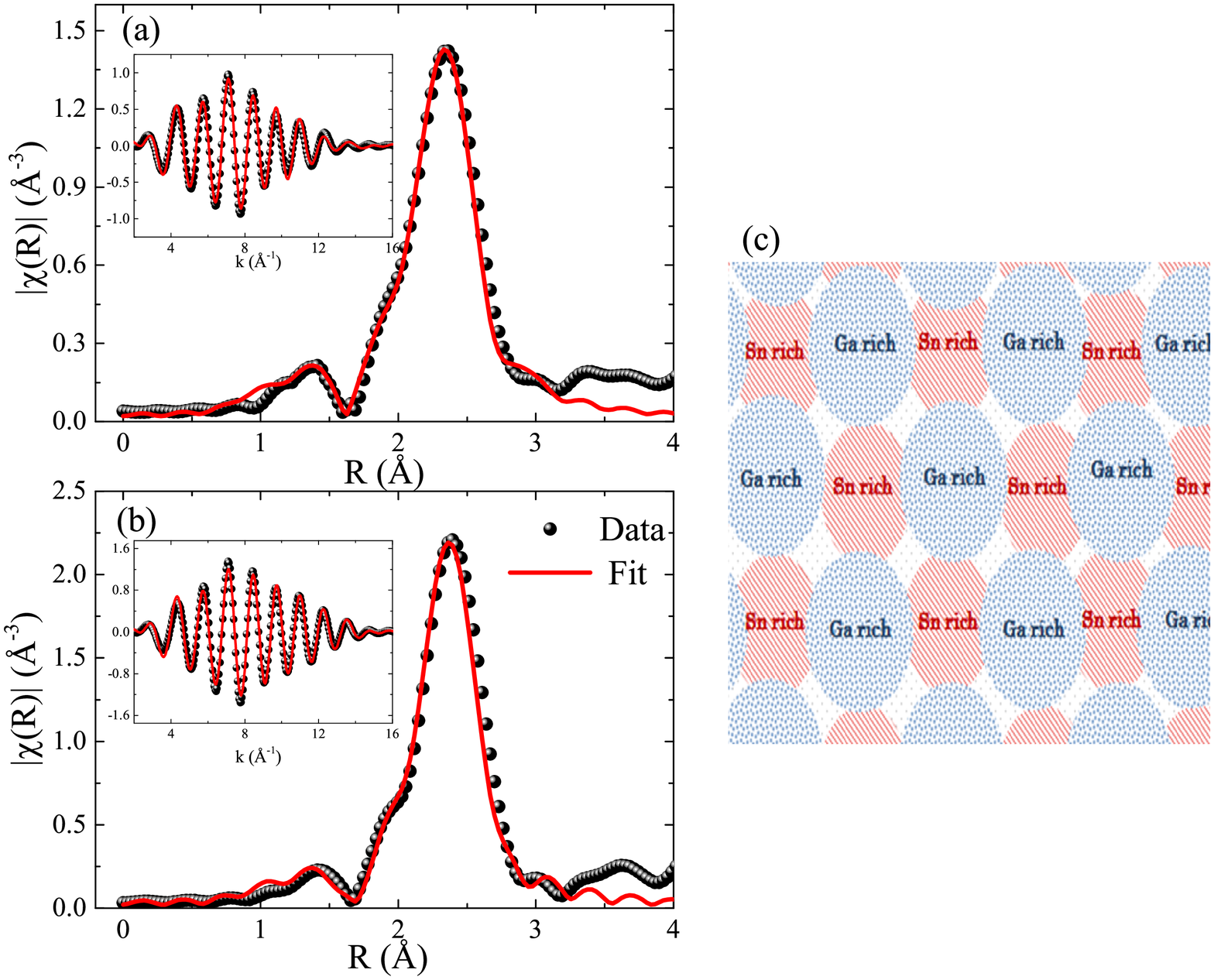}
\caption{The magnitude of Fourier transform of $ k^{2} $-weighted Mn K EXAFS spectra of Mn$_3$Ga$_{0.45}$Sn$_{0.55}$C recorded at (a) Low  temperature Mn K-edge and (b) Room temperature Mn K-edge along with corresponding best fits (solid lines). Inset shows the corresponding back transformed spectra of experimental data (solid spheres) and fit (solid line) in $k$ space. (c) Pictorial representation of the magnetic glassy state formed as a result of Ga rich and Sn rich regions in Mn$_3$Ga$_{0.45}$Sn$_{0.55}$C.}
\label{fig:exf}
\end{center}
\end{figure}

In order to obtain a more quantitative picture of the local structure around Mn in Mn$_3$Ga$_{0.45}$Sn$_{0.55}$C, the room temperature Mn K EXAFS data was fitted with correlations used for fitting EXAFS data of Mn$_3$GaC and Mn$_3$SnC in the R range of 1 to 3\AA ~ and $k$ range of 3 to 14 \AA$^{-1}$. It must be mentioned that at room temperature, the local structure around Mn in Mn$_3$GaC is as expected from the cubic symmetry \cite{Priolkar2016712} while the Mn$_6$C octahedra are distorted in Mn$_3$SnC \cite{Dias201548}. Constructing a model based on relative fractions of correlations in Mn$_3$GaC and Mn$_3$SnC as described earlier in Ref.  \onlinecite{Priolkar2016712} and \onlinecite{Dias201548} respectively and using the values of bond distances and $\sigma^2$ of various correlations obtained from fitting the EXAFS data of the two end members as first guess parameters resulted in a quite good fit as depicted in Figure \ref{fig:exf}(a). The relative fractions obtained were 0.52$\pm$0.08 of Mn$_3$GaC and 0.48$\pm$0.08 of Mn$_3$SnC.

At 100K (LT), the local cubic symmetry around Mn is broken and Mn-Mn and Mn-Ga bond distances are not equal \cite{Priolkar2016712}. Hence additional fitting parameters were introduced taking the number of fitting parameters to 13 as against 14 degrees of freedom. The number of fitting parameters were reduced to 11 by using same $\sigma^2$ for all Mn-C paths and a same value for all Mn-Mn paths. Such a fitting also resulted in a satisfactory fit (see Figure \ref{fig:exf}(b)) with relative fractions of Mn$_3$GaC and Mn$_3$SnC as 0.50$\pm$0.08 and 0.50$\pm$0.08 respectively. The parameter obtained from fitting RT and LT data are summarized in Table \ref{Table2}.

\begin{table}[]
\caption{Bond distance (R) and mean square disorder ($\sigma^2$) obtained from Mn K edge EXAFS analysis in Mn$_3$Ga$_{0.45}$Sn$_{0.55}$C recorded at 300K (RT) and 100K (LT). C.N. denotes the coordination number.}
\label{Table2}
\resizebox{\columnwidth}{!}{%
\begin{tabular}{cccccc}
\\
\hline
\multirow{2}{*}{Bond} & \multirow{2}{*}{C.N.} & \multicolumn{2}{c}{RT} & \multicolumn{2}{c}{LT} \\
 &  & R(\AA) & $ \sigma^{2} $ & R(\AA) & $ \sigma^{2} $ \\
\hline
\hline
&\multicolumn{4}{c}{Correlations of Mn$_3$GaC}&\\
Mn-C & 2 & 1.90(1) &0.005(1)& 1.94(3) & 0.003(1) \\
Mn-Mn & 8 & 2.70(1) &0.007(1)& 2.69(1) & 0.005(1) \\
Mn-Ga & 4 & 2.70(1) &0.003(2)& 2.71(1) & 0.002(1) \\
&\multicolumn{4}{c}{Correlations of Mn$_3$SnC}&\\
\multirow{2}{*}{Mn-C} & \multirow{2}{*}{2} & 1.91(1) & \multirow{2}{*}{0.005(1)}& 1.94(1) & \multirow{2}{*}{0.003(1)}\\
& & 2.07(1) & & 2.05(1) & \\
\multirow{2}{*}{Mn-Mn} & \multirow{2}{*}{8} & 2.71(1) & \multirow{2}{*}{0.007(1)} & 2.74(1) & \multirow{2}{*}{0.005(1)} \\
& & 2.93(1) & & 2.90(1) & \\
\multirow{2}{*}{Mn-Sn} & \multirow{2}{*}{4} & 2.93(1) & \multirow{2}{*}{0.004(1)}& 2.90(1) & \multirow{2}{*}{0.005(1)} \\
& & 2.71(1) & & 2.74(1) & \\
\hline
\end{tabular}
}
\end{table}

Thus the above results clearly indicate that though Mn$_3$Ga$_{0.45}$Sn$_{0.55}$C crystallizes in a structurally single phase, it consists of clusters that are either rich in Ga or rich in Sn distributed in a 54:46 ratio. The Mn atoms in the Ga rich clusters order antiferromagnetically with a propagation vector along $[\frac{1}{2}, \frac{1}{2}, \frac{1}{2}]$ direction as in case of Mn$_3$GaC and those Mn atoms in the Sn rich clusters order antiferromagnetically with propagation vector along $[\frac{1}{2}, \frac{1}{2}, 0]$ direction along with a net ferromagnetic moment. These clusters are large enough to present a magnetic Bragg peak in neutron diffraction but the interaction between two such clusters of the same type is severely limited by a cluster of the other type causing a slowing down of kinetics and resulting in a magnetically ``glassy" state. This clustered ground state is depicted pictorially in Figure \ref{fig:exf}c.

\section*{Conclusion}
In conclusion, Mn$_3$Ga$_{0.45}$Sn$_{0.55}$C though exhibits long range structural ordering, presents a non-ergodic ground state due to formation of Ga rich and Sn rich antiferromagnetically ordered clusters with different propagation vectors. The two types of clusters are distributed in a way that the magnetic interaction between two clusters of the same type is severely limited. Thus leading to a cluster glassy ground state. The main reason for such a cluster glass state is the difference in local structures of Mn atoms that find themselves in either Ga rich or Sn rich environment.

\section*{Acknowledgments}
The research work was supported by Board of Research in Nuclear Sciences (BRNS) under the project 2011/37P/06. Experimental assistance from M/s Devendra D. Buddhikot and Ganesh Jangam is gratefully acknowledged.

\bibliographystyle{apsrev4-1}
\bibliography{references}

\end{document}